\documentclass[aps,twocolumn,superscriptaddress,english,10pt,nofootinbib,preprintnumbers]{revtex4-2}

\usepackage{amssymb, amsmath, bm, dcolumn, epsf, graphicx, latexsym, slashed, simplewick}
\usepackage[utf8]{inputenc}
\usepackage[normalem]{ulem}

\usepackage{color}

\def\be{\begin{equation}}
\def\ee{\end{equation}}
\def\bea{\begin{eqnarray}}
\def\eea{\end{eqnarray}}

\usepackage{float}
\usepackage{hyperref}
\usepackage{comment}

\usepackage{xcolor}

\usepackage[normalem]{ulem}

\begin{document}

\preprint{MIT-CTP/5877}

\title{Kinetically Coupled Dark Matter Condensates}

\author{Michael W.~Toomey}
\affiliation{Center for Theoretical Physics -- a Leinweber Institute, Massachusetts Institute of Technology, Cambridge, MA 02139, USA}
\affiliation{ Department of Physics, Brown University, Providence, RI 02912-1843, USA}
\affiliation{ Brown Theoretical Physics Center, Brown University, Providence, RI 02912-1843, USA}

\author{Savvas M.~Koushiappas}
\affiliation{Center for Theoretical Physics -- a Leinweber Institute, Massachusetts Institute of Technology, Cambridge, MA 02139, USA}
\affiliation{ Department of Physics, Brown University, Providence, RI 02912-1843, USA}
\affiliation{ Brown Theoretical Physics Center, Brown University, Providence, RI 02912-1843, USA}

\author{Stephon Alexander}
\affiliation{ Department of Physics, Brown University, Providence, RI 02912-1843, USA}
\affiliation{ Brown Theoretical Physics Center, Brown University, Providence, RI 02912-1843, USA}

\begin{abstract}
Dark matter consisting of ultralight bosons can form a macroscopic Bose-Einstein condensate with distinctive observational signatures. While this possibility has been extensively studied for axions and axion-like particles---pseudoscalars with masses protected by shift symmetry---realistic models from string theory and other higher-dimensional theories predict more complex structures. Here we investigate a two-field generalization where an axion couples to a moduli field through its kinetic term, representing the phase and radial modes of a complex scalar field. We demonstrate that when this system forms a gravitationally bound Bose-Einstein condensate, the kinetic coupling produces dramatic modifications to cosmological evolution compared to the canonical single-field case. Most notably, the axion Jeans scale becomes dynamically dependent on the moduli field's evolution, fundamentally altering structure formation. By mapping existing observational constraints from canonical axion models to our two-field scenario, we identify regions of parameter space that are already excluded by current observations. In particular, consistency with observations requires that the moduli field must take on small field values, $\chi/M_{\rm pl} \ll 1$, throughout most of cosmic history for this class of axions to remain a viable description of all dark matter.
\end{abstract}

\maketitle

\section{Introduction}
The evidence supporting the existence of a  cold, dark, and pressure-less (or nearly so) matter in the Universe is firmly established in the last 50 years through a plethora of experimental measurements on a large range of spatial and temporal scales.  While the macroscopic behavior of dark matter is well understood, the micro-physical description still evades detection. The Large Hadron Collider (LHC) \cite{Aaboud:2019yqu,2017JHEP...10..073S}, together with direct and indirect detection experiments \cite{Drukier:1986tm,Goodman:1984dc,Akerib:2016vxi,Cui:2017nnn,Aprile:2018dbl,2020arXiv200304545F,2015JCAP...09..008F,2015PhRvD..91h3535G,2018ApJ...853..154A,2017PhRvD..95h2001A,2020Galax...8...25R,2016JCAP...02..039M,2017arXiv170508103I,2015arXiv150304858T,2018PhRvL.120o1301D, 2015ARNPS..65..485G,2020arXiv200610735K,2020arXiv200612488B} have placed strong constraints on the Weakly Interacting Massive Particle (WIMP), a well-motivated dark matter particle candidate motivated by physics at the weak scale  \cite{Steigman:1984ac}. 

One alternative dark matter candidate is the QCD axion \cite{Preskill:1982cy,Abbott:1982af,Dine:1982ah}. This dark matter candidate was originally proposed as a solution to the strong-CP problem in the Standard Model \cite{Peccei:1977hh,Wilczek:1977pj,Weinberg:1977ma}, where a new complex scalar field which is charged under a chiral, anomalous U(1) symmetry (the Peccei-Quinn symmetry U$(1)_{\rm PQ}$). The spontaneous breaking of this symmetry gives rise to a pseudo-Nambu-Goldstone boson, the QCD axion, which when added to the Standard Model can explain the anomalous neutron dipole moment. This hypothesis leads to a range of experimental predictions -- see \cite{DiLuzio:2020wdo} for a review of the current status of the QCD axion as a dark matter particle. 

Generally, axions fall under a broad category of light cold dark matter candidates. Most extensions to the Standard Model also predict other axion-like particles (ALP) —pseudoscalar spin-0 particles that arise from spontaneous breaking of global U(1) symmetries. These particles are non-relativistic in the present universe, making them suitable cold dark matter candidates. For the remainder of this paper, when we refer to axions we will use this term to encompass any generalized axion-like particle that originates from beyond the Standard Model theories.

As a dark matter candidate, one of the most interesting proposals in this class is for a very light, typically $m \sim  \mathcal{O}(10^{-22}~{\rm eV})$, axion with wave-like properties that form a Bose-Einstein Condensate (BEC) \cite{Turner:1983he,Press:1989id,Sin:1992bg,Hu:2000ke,Goodman:2000tg,Peebles:2000yy,Amendola:2005ad,Schive:2014dra,Chavanis:2011zi,Chavanis:2011zm,Alexander:2016glq,Alexander:2018fjp,Alexander:2024nvi} -- for  reviews see~\cite{Hui:2016ltb,Ferreira:2020fam}. Assuming a potential  $V(\phi) \sim m^2 F^2$, where $F$ is bounded by  $M_{\mathrm{planck}} < F < \Lambda_{\mathrm{GUT}}$, the abundance of such dark matter candidate is of order the observed dark matter abundance \cite{Hui:2016ltb}
$
\Omega_a \sim 0.1 (F / 10^{17})^2 (m / 10^{-22} {\mathrm{eV}})^{1/2} 
$. 
Such a candidate has well understood, but unique phenomology including wave-like properties\cite{Capanelli:2025nrj}, the formation of topological substructure \cite{Rindler-Daller:2011afd,Alexander:2019puy,Alexander:2021zhx,Alexander:2019qsh} and a de~Broglie wavelength that is of Galactic scales with soliton formation at the center of dark matter halos \cite{Chavanis:2011zi,Chavanis:2011zm}.

Recently, however, it has been emphasized that the most general string axions exhibit a coupling between moduli fields and the kinetic term for the axion \cite{Alexander:2022own,Bernardo:2022ztc,Brax:2023qyp}. While present in string theory, the origin of such kinetic couplings fundamentally stem from the existence of extra-dimensions and is therefore likely to be a feature of other theories beyond the Standard Model which rely on compactification -- though we will focus here on a string inspired model. Despite their theoretical ubiquity, the phenomenology of these couplings has not been explored extensively, though some work has shown the potential for possibly alleviating tensions in the concordance model \cite{Alexander:2022own}. 

In this work we push the understanding of kinetically coupled axions further by studying the modified dynamics for gravitationally bound axion Bose-Einstein condensates with a kinetic coupling to a moduli field. We find that this coupling enhances the Jeans scale by factors that depend on the moduli field value, dramatically altering structure formation on small scales. By mapping existing astrophysical constraints into our framework, we show that viable parameter space requires the moduli field to remain at small values throughout cosmic history—a constraint that has profound implications if this field also serves as a form of quintessence. Our analysis reveals that kinetic couplings, far from being negligible corrections, fundamentally alter the phenomenology of ultralight dark matter and may forge an unexpected connection between the nature of dark matter and dark energy. This is particularly interesting in light of recent DESI measurements \cite{DESI:2025zgx,DESI:2024mwx} and the possibility for apparent phantom behavior in two-field models similar to what is studied here \cite{Smith:2024ibv}.

The remainder of this paper is organized as follows: Section~\ref{KCUP} introduces the kinetically coupled model and derives the modified Gross-Pitaevskii equation, Section~\ref{Probes} examines observational constraints, Section~\ref{Theory_Imp} discusses theoretical implications, and Section~\ref{DNC} concludes with a summary and future directions.

\section{Kinetically Coupled Condensates} 
\label{KCUP}

The form of the most general action for axions that arise as the result of compactification of higher dimensions is  \cite{Alexander:2022own,Bernardo:2022ztc,Burgess:2021obw,Smith:2024ibv},
\begin{equation}
    S = \int d^4x  \sqrt{-g} \left\{- \frac{1}{2}  (\partial_\mu \chi)^2 - \frac{1}{2} (\partial_\mu \phi)^2  f- V(\chi, \phi)\right\},
    \label{action}
\end{equation}
where $f =f(\chi) = e^{\lambda \chi}$.\footnote{We simply refer to this as $f$ elsewhere in the text and should not be confused with $f_\phi$ the axion decay constant.} This realizes a two-field model where the axion $\phi$ is coupled to a moduli-field $\chi$ that is minimally coupled to gravity. Nominally, in string theory, $\lambda$ is an $\mathcal{O}(1)$ value corresponding to ones choice of string compactification. Here $\chi$ is endowed with a potential of the form $V(\chi) = A e^{-\alpha \chi}$ and the axion has a potential of the form $V(\phi) = m^2f_\phi^2 [ 1 - \cos(\phi/f_\phi)]$ where $m$ and $f_\phi$ are the axion mass and decay constant, respectively. 

We adopt a Friedmann-Robertson-Walker metric to study the cosmological evolution in this model. To  understand the formation of a Bose-Einstein condensate, we take the non-relativistic limit of the theory.\footnote{For a discussion of the more general case, see \cite{Guth:2014hsa}.} Note that when we do so, we treat $\chi$ as a background field such that it does not contribute to the quantum behavior of the condensate. First we expand the potential to fourth order and isolate terms for the $\phi$-field,
\begin{equation}
    S = \int d^4x \sqrt{-g} \left[  -\frac{1}{2} (\partial_\mu \phi)^2 f - \frac{1}{2} m^2 \phi^2 - \frac{g_\phi}{4!} \phi^4 \right],
\end{equation}
with $g_\phi = m^2/f_\phi^2$ the self-coupling constant. To calculate the  behavior of the axion once it forms a Bose-Einstein condensate, we can take the non-relativistic limit of the Lagrangian by first writing $\phi$ in terms of a complex scalar field $\psi$ where
\begin{equation}
    \phi(t,x) = \frac{1}{\sqrt{2m a^3}}\left[e^{-imt}\psi(t,x) + h.c.\right]
    \label{nr-limit}
\end{equation}
and which has the corresponding conjugate momenta, $\Pi(t,x)$. Dropping the fast oscillating terms which are proportional to $e^{\pm imt}$, the non-relativistic Hamiltonian density is found to be,
\begin{equation}
\begin{split}
    \hat{\mathcal{H}} &= a(t)^3 \left\{\frac{f}{2m a^2}\nabla \hat{\psi}\nabla \hat{\psi}^\dagger + \frac{1}{2} m   \hat{\psi^\dagger}\hat{\psi} (f - 1) \right. \\ &+ \left. \frac{g_\phi}{16 m^2} \hat{\psi^\dagger}\hat{\psi^\dagger}\hat{\psi}\hat{\psi} +  V(\chi) \right\}  + \hat{\mathcal{H}}_{\rm grav}.
    \label{hamden}
\end{split}
\end{equation}
Here, $\hat{\mathcal{H}}_{\rm grav}$ represents the Hamiltonian contribution from gravity. It can be found by first noting that the Bose-Einstein number density is given by,
\begin{equation}
    \hat{n}(x) = \hat{\psi(x)}\hat{\psi}^\dagger(x),
\end{equation}
and therefore the mass density is,
\begin{equation}
    \hat{\rho}(x) = m \hat{\psi}^\dagger(x) \hat{\psi}(x). 
\end{equation}
This implies the gravitational contribution to the Hamiltonian is,
\begin{equation}
    \hat{\mathcal{H}}_{\rm grav} = -\frac{Gm^2}{2} \int d^3x' \frac{\hat{\psi^\dagger}(x')\hat{\psi^\dagger}(x)\hat{\psi}(x')\hat{\psi}(x)}{\left| x - x' \right|}.
\end{equation}
which explicitly completes the full Hamiltonian. 

From the Heisenberg equation of motion, the modified Gross–Pitaevskii equation for $\psi$ is 
\begin{equation}
    \begin{split}
    i \psi' = &-\frac{i}{2} \left(3 a H + \frac{ d \ln f}{d \eta}  \right)\psi -\frac{1}{2m a}\nabla^2 \psi  \\ &+ \frac{ma}{2}\left(1 - \frac{1}{f}\right)\psi + \frac{a g_\phi}{8m^2 f} \left|\psi\right|^2 \psi + \frac{m a}{f} \psi \Phi,
\end{split}
    \label{frwgross}
\end{equation}
where $'$ denotes a derivative w.r.t to conformal time and $H$ is the Hubble parameter, which together with the Poisson equation
\begin{equation}
    \nabla^2 \Phi = 4\pi G a^2 \left( m|\psi|^2 - \bar\rho \right),
\end{equation}
and Klein-Gordon equation for $\chi$ 
\begin{equation}
    {\chi''} + 2 a H {\chi'} - \frac{1}{2}f_\chi {\phi'}^2 + a^2V_\chi = 0,
\end{equation}
forms the system of equations that govern the dynamics of the $\psi$ and $\chi$ fields.  

As anticipated, there are modifications from the standard form of the Gross-Pitaevskii equation (see also~\cite{Guth:2014hsa}). In particular, there is an extra friction term that can either enhance or suppress Hubble friction depending on the sign of $d \ln f / d\eta$. In addition, the axion and gravitational potentials are suppressed by a factor of $f$. This suppression means that the contribution from the mass term that is typically canceled by an equal but opposite contribution from the axion kinetic term is now present -- see first term on the second line of Eq.~\ref{frwgross}. 

Note that in Eq.~\ref{frwgross} we have dropped all the hats on $\psi$. This comes from a mean field approximation where we decompose $\hat{\psi}$ into a background piece $\psi$ (the expectation value of $\hat{\psi}$) and fluctuations $\delta \hat{\psi}$. In the limit of high occupancy number, $\mathcal{N}$, $\hat{\psi} \sim \psi$ where $\psi$ is now a c-number -- i.e. since $\delta \hat{\psi}/\psi \sim 1 / \sqrt{\mathcal{N}}$.

\begin{figure}[!t]
    \centering
    \includegraphics[width=\linewidth]{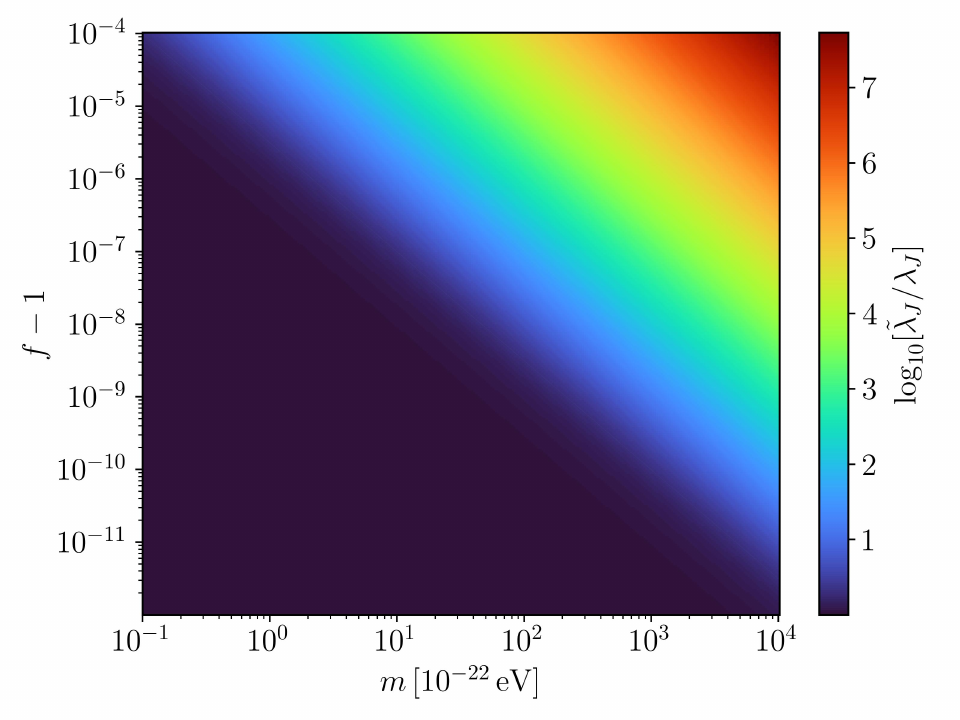}
    \caption{Enhancement in the physical size of kinetically coupled soliton cores as a function of axion mass and the quantity $f-1$. Color corresponds to the logarithm of the ratio between the kinetically coupled Jeans length from Eq.~\ref{jeans} to the standard Jeans length of Eq.~\ref{eq:standard_kJ} at the redshift of recombination.}
    \label{fig:stab}
\end{figure}

To gain physical insight into the modified dynamics, it is instructive to rewrite the Gross-Pitaevskii equation as a system of fluid equations.  We begin by expressing the wave function as a modulus and phase,
\begin{equation}
    \psi(r,\eta) = \sqrt{\frac{\rho(r,\eta)}{m}} e^{i\theta(r,\eta)}.
\end{equation}
After plugging into Eq.~\ref{frwgross} we find the Madelung equations,
\begin{equation}
    \frac{d \rho}{d \eta} + 3 a H \rho + {\bm{\nabla}} \cdot \left( \rho {\bf{v}}\right) = - \frac{ d \ln f}{d \eta} \rho,
    \label{22}
\end{equation}
\begin{equation}
    \frac{d {\bf{v}}}{d\eta} + a H {\bf{v}} + \left( {\bf{v}} \cdot {\bm{\nabla}} \right) {\bf{v}} = - \frac{{\bm{\nabla}}\Phi}{f}  - \frac{P_{\rm Int}}{f} +  P_{\rm Q},
    \label{euler2}
\end{equation}
where 
\begin{equation}
P_{\rm Q} = \frac{1}{2m^2a^2} {\bm{\nabla}}  \left(\frac{\nabla^2 \sqrt{\rho}}{\sqrt{\rho}} \right)  \nonumber
\end{equation}
is the usual quantum pressure,  $P_{\rm Int}$ is the pressure contribution from the interaction term
\begin{equation}
    P_{\rm Int} = \frac{g_\phi}{8 m^4}{\bm{\nabla}}\rho,
\end{equation}
and ${\bf{v}} \equiv {\bm{\nabla}}\theta / m a$. One of the clear consequences of the kinetic coupling in this model is a shift in the relative strength of the quantum pressure to other sources of ``pressure.'' From the r.h.s. of Eq.~\ref{euler2}, we see that in the hydrostatic limit there is a suppression by a factor of $f$,
\begin{equation}
    \left| \frac{P_{\rm Q}}{\bm{\nabla}\Phi + P_{\rm Int} + \ldots}  \right | = f.
    \label{qrat}
\end{equation}
In the case of string axions specifically, the self-interaction and other higher-order terms are strongly suppressed relative to the quadratic as the field undergoes harmonic oscillations, given  $\phi \ll f_\phi\sim M_{pl}$. Therefore in what follows, we do not consider the effects of the self-interactions and all results below are only applicable to the case of string axions with no self-coupling.   In this limit, the model is then qualitatively similar to screening the axion's mass by a factor of $\sqrt{f}$.  From this result it is evident  that a direct consequence of such a coupling is the possibility for {\it an enhanced size of the central soliton in dark matter halos from a relative enhancement in the quantum pressure}. 

The presence of the kinetic coupling will result in a modification to the growth of perturbations in the condensate. To get a more quantitative understanding of this change we can perturb the Gross–Pitaevskii equation, Eq.~\ref{frwgross}, and derive the modified Jeans scale $\Tilde{k}_J$ (corresponding to the physical Jeans length  $\tilde{\lambda}_J = 2\pi/\tilde{k}_J$). Doing so we find that,
\begin{equation}
    \left(\frac{\Tilde{k}_J}{a}\right)^2 = \sqrt{{\cal{F}}^2+ \frac{6 H^2 m^2 \Omega_\phi}{ f}} - {\cal{F}}
    \label{jeans}
\end{equation}
where $\Omega_\phi\equiv\rho_\phi/\rho_{\rm crit}$ is the axion density and $\mathcal{F}$ is given by
\begin{equation} 
{\cal{F}} \equiv \frac{m^2(f - 1)}{2f}.
\end{equation}
Note that in the limit as $f \rightarrow 1$ the modified Jeans scale reduces to the standard result, 
\begin{equation} 
\lim_{f\to 1} \left(\frac{\tilde{k}_J}{a}\right)^2 \rightarrow  \sqrt{6 H^2 m^2 \Omega_\phi } = \left( \frac{k_J}{a} \right)^2 = \left( \frac{2 \pi}{a \, \lambda_J } \right)^2.
\label{eq:standard_kJ}
\end{equation} 

In Fig.~\ref{fig:stab} we show  the ratio of the modified Jeans scale to the standard scale as a function of the axion mass and $f - 1$. Note that here we have assumed that $\chi^\prime$ is negligible, but its easy to restore this effect by replacing $H \rightarrow H + (3 a)^{-1} d\log{f}/d\eta$ -- as in a scenario where $\chi$ plays the role of a quintessence-like field \cite{Caldwell:1997ii}.

Note that as ${\cal{F}} \sim m^2$, heavier axions are more impacted by the effects of the kinetic coupling. This can be understood from Eq.~\ref{jeans} as a balance between the standard contribution and new terms. For example, the new terms depend more strongly on the axion mass, e.g. the ratio of the two innermost terms goes as $m^2$. So for a fixed $f$, axions with larger masses incur larger deviations from the canonical axion. On the contrary, smaller masses suppress the first and third terms of Eq.~\ref{jeans} and requires correspondingly larger values for $f$ to appreciably modify the Jeans scale. 

\begin{figure}[!t]
    \centering
    \includegraphics[width=\linewidth]{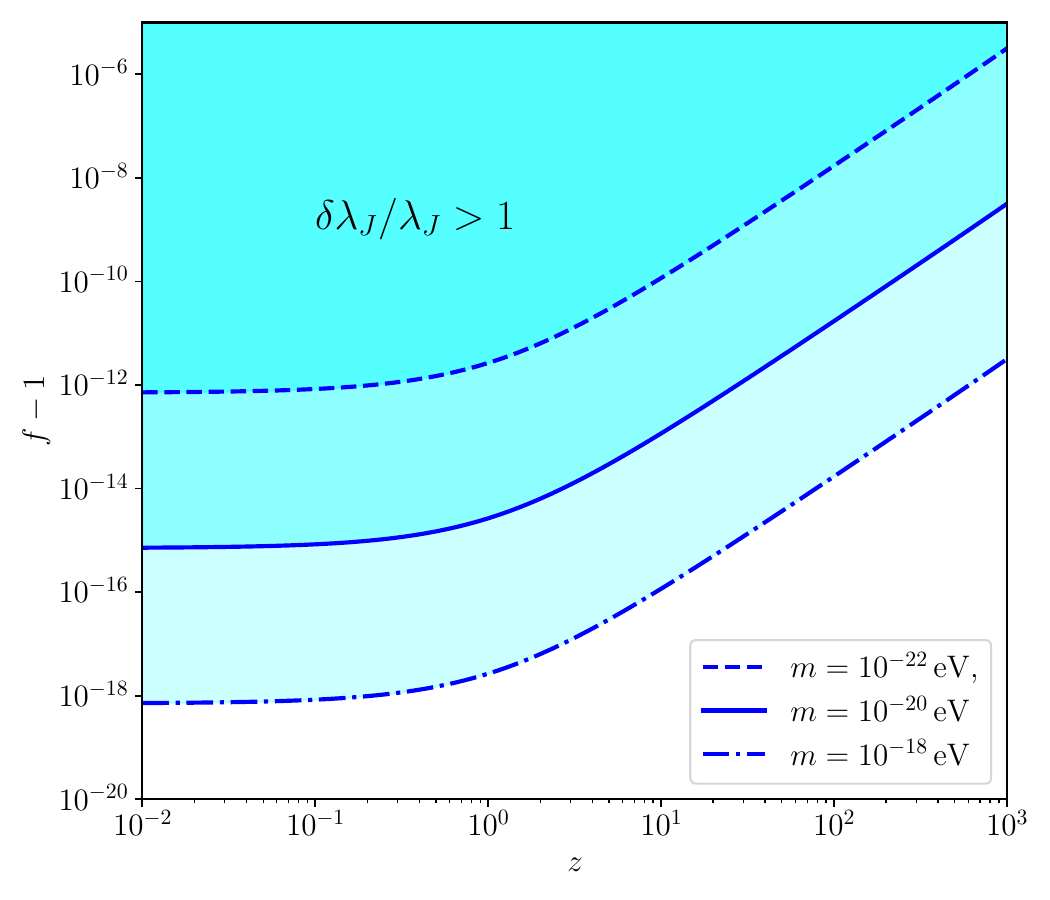}
    \caption{Sensitivity of Jeans scale enhancement as a function of redshift and kinetic mixing strength $f $ for three fiducial axion masses. The region above each curve results in a larger de~Broglie wavelength, with $\delta \lambda_J / \lambda_J >1$ -- see text for details. }
    \label{fig:jeans_ev}
\end{figure}

Figure~\ref{fig:jeans_ev} shows the regions where $\delta \lambda_J/\lambda_J  = (\tilde{\lambda}_J/\lambda_J - 1) > 1$ as a function of redshift for three fiducial axion masses ($\tilde{\lambda}_J/\lambda_J$ is always $\tilde{\lambda}_J/\lambda_J \ge 1$). As is evident, effects of the kinetic coupling is suppressed in the early-Universe relative to late-times.

\section{Experimental Probes}
\label{Probes}

As we just showed, the coupling between the two fields $\{ \phi, \chi\}$  results in a dynamical Jeans scale that depends on the evolution of the $\chi$-field (and correspondingly $f$). This has wide ranging implications as a change in the Jeans length will have consequences for a multitude of observables. 
Here, we address some of these observational consequences of the unique dynamics of this model.

Similar to the standard axion scenario, at the level of the matter power spectrum this model will produce a step, i.e. a suppression, for $k > \Tilde{k}_J$ given that the scalar perturbations are oscillatory below $\tilde{k}_J$ \cite{Marsh:2010wq,Amendola:2005ad,Allali:2021azp}. This deviation can be approximated as 
\begin{equation}
    \frac{\mathcal{P}_{\phi + {\rm cdm}}}{\mathcal{P}_{\rm cdm}} \approx \left( \frac{\Tilde{k}^{\rm eq}_J}{k}\right)^{8(1 - q)}
\end{equation}
where $q = (\sqrt{25 - 24\mathcal{G} } - 1)/4$ and $\mathcal{G}$ is the fraction of dark matter in the axion \cite{Marsh:2010wq, Amendola:2005ad}. In principle, when all other parameters remain constant, increasing $f$ both strengthens the suppression of structure and extends that suppression to larger scales. Observationally, this suppression can result in a noticeable decrease in the number of the lightest halos and subhalos and generic expectations from hierarchical structure formation.

This effect can be seen elsewhere. For example, from Eq.~\ref{22}, it is easy to see that the $\chi$ dependence roughly affects the condensate density as
\begin{equation}
    \rho_{\rm \phi} \propto a^{-3} e^{\lambda \Delta\chi},
\end{equation}
where $\Delta \chi$ is taken to be the difference in field value over the period of interest. Given the naive theory expectation that $\lambda$ should be no more than $\mathcal{O}(1)$, this implies that the evolution of $\chi$ can \textit{suppress} the BEC dark matter density by factors of 10 - 100 for $|\Delta \chi| \approx 1$. This has the potential to result in  a dramatic departure from the standard cold, pressureless dark matter and would have further ramifications for the growth of structure. For example, the presence of such a coupling to dark matter could make an appreciable change to the amplitude of the cosmic microwave background  at large-scales through an enhancement of the late integrated Sachs-Wolfe (ISW) effect, 
\begin{equation}
    \label{eq:lISW}
    C^{\rm ISW}_\ell \propto \int \frac{d k}{k^2} P(k) D \left[ \int d\eta D \left( \frac{D^\prime}{D} - \mathcal{H} + \frac{d \ln{f}}{d\eta}  \right) j_\ell (k\eta) \right]^2. 
\end{equation}
Here,  $D$ the growth factor, $P(k)$ the power spectrum,  $j_\ell (k\eta)$ is the spherical Bessel function, and $\mathcal{H} = aH$ is the conformal Hubble parameter. Here we see the implicit dependence on the field velocity of $\chi$ via $f$. Nominally, the late-ISW effect is minimal during periods of matter domination as the growth factor goes as $D \sim a$ such that $\Phi' \approx0$. However, if $\left. d \ln{f} \right/d\eta  \sim\lambda \chi^\prime/\mathcal{H}$ becomes large enough we could anticipate a non-negligible ISW contribution to the CMB.  Note also that beyond the $\chi$ dependence in Eq.~\ref{eq:lISW}, there is also nontrivial $\chi$ dependence for both $\mathcal{H}$, $D$, and $\eta$ at the level of their equations of motion.

\section{Theory Implications}
\label{Theory_Imp}

\begin{figure}[!t]
    \centering
    \includegraphics[width=\linewidth]{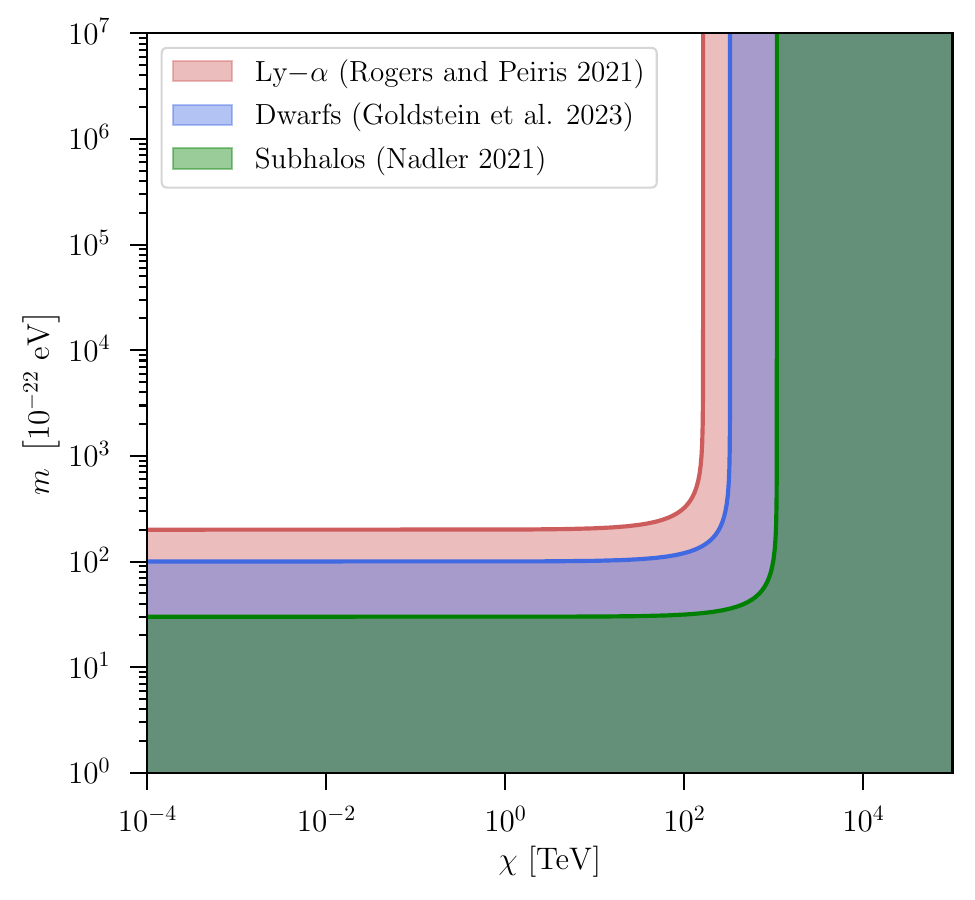}
    \caption{Inferred constraints on the kinetically coupled axion parameter space at $z = 0$ from Ly-$\alpha$ \cite{Rogers:2020ltq} (red), dwarf galaxies \cite{Goldstein:2022pxu} (blue), and subhalo counts \cite{DES:2020fxi} (green). The constrained regions of parameter space are established by enforcing that the modified Jeans scale, Eq.~\ref{jeans}, does not exceed that of a standard axion, Eq.~\ref{eq:standard_kJ}, for the respective constraints. Notice that such a coupling strongly constrains the allowed values of moduli field $\chi$ where we have assumed $\lambda = 1~ M^{-1}_{\rm pl}$, consistent with the $\mathcal{O}(1)$ theory expectations.}
    \label{fig:chi_constraint}
\end{figure}

Given the modified Jeans scale in this model,  we can reinterpret existing astrophysical constraints on fuzzy dark matter as bounds on the kinetic coupling. In particular, we take observational lower limits on the axion mass—derived under the assumption of a canonically normalized kinetic term—and require that the Jeans scale in our model does not exceed that of a canonical axion with mass $m_X$ associated with each constraint. That is, for each probe $X$, we enforce:
\begin{equation}
    \tilde{\lambda}_J(m, \chi) \leq \lambda_J(m_X),
\end{equation}
where $\tilde{\lambda}_J$ is the modified Jeans length from Eq.~\ref{jeans}, and $\lambda_J(m_X)$ is the standard Jeans scale associated with mass $m_X$. This condition defines an exclusion region in the $(m, \chi)$ parameter space, which we display in Fig.~\ref{fig:chi_constraint}, using constraints from Lyman-$\alpha$ forest data \cite{Rogers:2020ltq}, dwarf galaxy kinematics~\cite{Goldstein:2022pxu}, and subhalo counts \cite{DES:2020fxi}. These bounds are derived under the assumption that axions make up all of the dark matter.

In our model, the axion kinetic term is rescaled by a field-dependent function $f(\chi) = e^{\lambda \chi}$, a form motivated by dimensional reduction in string theory. This field $\chi$ typically plays the role of a geometric modulus, controlling the volume of compactified extra dimensions as $\mathcal{V} \sim e^{\gamma \chi}$. As $\chi$ increases, the effective quantum pressure in the Gross–Pitaevskii equation increases, leading to a growth in the Jeans scale and enhanced suppression of structure on small scales. As a result, astrophysical probes place \emph{upper bounds} on the allowed values of $\chi$, with the tightest constraints at lighter axion masses. The cosmological constraints significantly restrict the viable region of parameter space. As shown in Fig.~\ref{fig:chi_constraint}, current bounds already exclude large values of $\chi$ for the whole fuzzy dark matter mass range. This parameter space is expected to be further constrained with future 21cm and Ly-$\alpha$ measurements combined with the recent developments in the effective field theory description for these bias tracers \cite{Ivanov:2023yla,Ivanov:2024jtl,Qin:2022xho}.

A striking feature of our results is the universality of the constraint on the moduli field value seen in Fig.~\ref{fig:chi_constraint}. Across the fuzzy dark matter mass range, cosmological observations require $\chi \lesssim 10^2$ TeV. This represents a profound statement about the viability of kinetically coupled axions from string theory: any such model must either have extremely small moduli field values ($\chi/M_{\rm pl} \lesssim 10^{-14}$) over a majority of cosmic history or invoke additional mechanisms to suppress the coupling. With the previous caveat that such effects are strongly suppressed before recombination as shown in Fig.~\ref{fig:jeans_ev}. This constraint is particularly severe given theoretical expectations. In typical string compactifications, moduli fields naturally explore field ranges of order the Planck scale during cosmological evolution. Our bound of $\chi \lesssim 10^2$ TeV implies that either: (i) the relevant moduli must be strongly stabilized near the origin of field space, (ii) the coupling $\lambda$ must be significantly smaller than the naive $\mathcal{O}(1/M_{\rm pl})$ estimate, or (iii) if $\chi$ does undergo a large field excursion it must do so before recombination. This tension between theoretical expectations and observational constraints provides a concrete example of how precision cosmology is in a unique position to constrain the string landscape.

\section{Discussion and Conclusion}
\label{DNC}

In this work we have studied the basic dynamics of a generalization of the standard ultra-light axion dark matter model. In particular, we have focused on a two-field model where the axion's kinetic term acquires a field-dependent normalization through coupling to a moduli field $\chi$. This form, with $f = e^{\lambda \chi}$, represents the generic structure expected when axions arise from dimensional reduction in string theory. Provided that this moduli field is not the dilaton—ensuring it is only minimally coupled to gravity—the model avoids fifth force constraints while capturing essential features of string-inspired axions. We stress however, that these types of couplings are expected features of any models that rely on compactification.

The modified Gross-Pitaevskii equation (Eq.~\ref{frwgross}) reveals several important modifications to condensate dynamics. The gravitational potential and self-interactions all acquire a suppression factor $f^{-1}$, while an additional friction term proportional to $d \ln f/d\eta$ appears that can either augment or suppress the friction that arises from the expansion of the Universe. These changes fundamentally alter the balance of forces determining the structure of dark matter, namely the size and density of the resulting soliton core. 
Our key finding is that the kinetic coupling can enhance the Jeans scale by factors of 2-10 for phenomenologically relevant parameters. This enhancement is most pronounced for heavier axions ($m > 10^{-21}$ eV), where the modified  quantum pressure can dramatically alter structure formation. By translating existing astrophysical constraints into our framework, we have found stringent bounds on the late-time moduli field value,  $\chi \lesssim 10^{2}~{\rm TeV}$.

This universal bound on $\chi$ across all viable axion masses places severe constraints on the types of quintessence models compatible with kinetically coupled fuzzy dark matter. The maximum field excursion corresponds to $\chi/M_{\rm pl} \lesssim 10^{-14}$ in the late-Universe, effectively requiring the moduli field to remain a tiny fraction of the Planck scale throughout post-recombination history. Such small field values are common in quintessence models with extremely flat potentials near the origin. Note that this effectively rules out large-field quintessence models kinetically coupled to axion dark matter, suggesting that if such couplings exist in nature, dark energy may necessarily arise from small-field dynamics—a conclusion that resonates with the swampland conjectures regarding scalar field excursions in quantum gravity (although it is of course possible that the associated moduli field does not play the role of dark energy, or is a minor contribution). 

Beyond these technical considerations, our framework opens an intriguing possibility: the potential for a unified description of the dark sector. Indeed, as pointed out in \cite{Brax:2023qyp,Smith:2024ibv}, models that exhibit such non-canonical kinetic terms have the potential to result in interesting observables in light of recent results from DESI DR2 BAO measurements. Specifically, the effective equation of state for dark energy can appear phantom, in our case
\begin{equation}
    w_{\chi,\rm eff} = \frac{\chi'^2 - 2a^2V(\chi) + a^2(1 - f^{-1})\rho_\phi}{\chi'^2 + 2a^2V(\chi) + a^2(1 - f^{-1})\rho_\phi},
\end{equation}
which results in $w_{\chi,\rm eff} < -1$ when $f < 1$.\footnote{The coupling between these fields prevents us from expressing their respective densities and pressures as independent quantities. This effective dark energy equation of state emerges when we isolate the dark energy component by subtracting the axion's dark matter density contribution from the coupled system's total density and pressure.}  Nominally, this requires a negative $\lambda$ that goes against typical string theory expectations \cite{Dine:1985he}, but which is possible to construct if $\lambda \rightarrow \lambda(\chi) \approx \lambda - \mathcal{O}(\chi)$, see, for example, \cite{Alexander:2019rsc}. Note that in this scenario, nothing is truly phantom, rather dark energy can appear to be phantom due precisely to the coupling between the dark sectors —a feature of particular interest given DESI's preliminary evidence for dynamical dark energy.\footnote{See \cite{Wang:2025vfb} and \cite{Efstathiou:2025tie} for alternative perspectives on the evidence of evolving dark energy from DESI data.} This connection suggests that searches for dark energy evolution might simultaneously constrain the microscopic nature of dark matter. Note also that this difference in sign for $\lambda$ inverts the behavior that we have focused on in this work: in this scenario the coupling would work to \textit{reduce} the Jeans scale.

Several important caveats shape the interpretation of our results. All constraints presented assume axions constitute the entirety of the dark matter; if axions comprise only a fraction of the total density, our bounds would weaken accordingly as structure formation could proceed closer to that of the standard cold dark matter scenario. Additionally, our results depend sensitively on the specific form of $f$. While the exponential coupling is well-motivated from string theory, more general functional forms are possible, and the dimensionless combination $\lambda M_{\rm pl}$ could deviate from our assumed $\mathcal{O}(1)$ value in specific compactifications.

Finally, we note that in principle, axions could also couple to the moduli field through the potential rather than the kinetic term, as considered in models such as \cite{Alexander:2019rsc}. In that case, the suppression of the gravitational and self-interaction potentials would be absent, but an extra friction term proportional to $d \ln f / d \eta$ would remain. This would still impact the condensate evolution and the Jeans scale, depending on the sign and magnitude of the field velocity $\chi'$. At the level of the fluid equations, this corresponds to preserving the $\lambda \chi'$ term in the continuity equation while reverting the Euler equation to its standard form.

The kinetic coupling examined here represents a generic feature of higher-dimensional theories containing axions, rather than a fine-tuned scenario. As observational probes of small-scale structure achieve increasing precision—ranging from dwarf galaxy kinematics~\cite{Goldstein:2025lvl} to 21-cm intensity mapping~\cite{Weltman:2018zrl}—and as surveys such as DESI~\cite{DESI:2025zpo} constrain the expansion history to percent-level accuracy~\cite{Moresco:2022phi}, deviations from the predictions of minimal dark matter models may become detectable. Our analysis demonstrates that even modest couplings between dark matter and moduli fields can produce novel observable signatures. These findings suggest that systematic searches for such signatures in upcoming datasets could either validate or rule out broad classes of string-theoretic dark matter models, potentially resolving long-standing questions about the microphysical nature of dark matter.

\section{Acknowledgements}
We thank Heliudson Bernardo, Mark Hertzberg, and Austin Joyce for useful discussions. This material is based upon work supported by the U.S. Department of Energy, Office of Science, Office of High Energy Physics of U.S. Department of Energy under grant Contract Number  DE-SC0012567. M.~W.~T.  acknowledges financial support from the Simons Foundation (Grant Number 929255). S.~M.~K. is  supported by National Science Foundation (NSF) No. PHY-2412666.

\bibliography{bibo}

\begin{thebibliography}{10}

\bibitem{Aaboud:2019yqu}
Morad Aaboud et~al.
\newblock {Constraints on mediator-based dark matter and scalar dark energy models using $\sqrt s = 13$ TeV $pp$ collision data collected by the ATLAS detector}.
\newblock {\em JHEP}, 05:142, 2019.

\bibitem{2017JHEP...10..073S}
Albert~M Sirunyan et~al.
\newblock {Search for new physics in the monophoton final state in proton-proton collisions at $ \sqrt{s}=13 $ TeV}.
\newblock {\em JHEP}, 10:073, 2017.

\bibitem{Drukier:1986tm}
A.~K. Drukier, Katherine Freese, and D.~N. Spergel.
\newblock {Detecting Cold Dark Matter Candidates}.
\newblock {\em Phys. Rev.}, D33:3495--3508, 1986.

\bibitem{Goodman:1984dc}
Mark~W. Goodman and Edward Witten.
\newblock {Detectability of Certain Dark Matter Candidates}.
\newblock {\em Phys. Rev.}, D31:3059, 1985.

\bibitem{Akerib:2016vxi}
D.~S. Akerib et~al.
\newblock {Results from a search for dark matter in the complete LUX exposure}.
\newblock {\em Phys. Rev. Lett.}, 118(2):021303, 2017.

\bibitem{Cui:2017nnn}
Xiangyi Cui et~al.
\newblock {Dark Matter Results From 54-Ton-Day Exposure of PandaX-II Experiment}.
\newblock {\em Phys. Rev. Lett.}, 119(18):181302, 2017.

\bibitem{Aprile:2018dbl}
E.~Aprile et~al.
\newblock {Dark Matter Search Results from a One Ton-Year Exposure of XENON1T}.
\newblock {\em Phys. Rev. Lett.}, 121(11):111302, 2018.

\bibitem{2020arXiv200304545F}
Francis {Froborg} and Alan~R {Duffy}.
\newblock {Annual Modulation in Direct Dark Matter Searches}.
\newblock {\em arXiv e-prints}, page arXiv:2003.04545, March 2020.

\bibitem{2015JCAP...09..008F}
{Fermi LAT Collaboration}.
\newblock {Limits on dark matter annihilation signals from the Fermi LAT 4-year measurement of the isotropic gamma-ray background}.
\newblock {\em \jcap}, 2015(9):008, September 2015.

\bibitem{2015PhRvD..91h3535G}
Alex {Geringer-Sameth}, Savvas~M. {Koushiappas}, and Matthew~G. {Walker}.
\newblock {Comprehensive search for dark matter annihilation in dwarf galaxies}.
\newblock {\em \prd}, 91(8):083535, April 2015.

\bibitem{2018ApJ...853..154A}
A.~Albert et~al.
\newblock {Dark Matter Limits From Dwarf Spheroidal Galaxies with The HAWC Gamma-Ray Observatory}.
\newblock {\em Astrophys. J.}, 853(2):154, 2018.

\bibitem{2017PhRvD..95h2001A}
S.~Archambault et~al.
\newblock {Dark Matter Constraints from a Joint Analysis of Dwarf Spheroidal Galaxy Observations with VERITAS}.
\newblock {\em Phys. Rev. D}, 95(8):082001, 2017.

\bibitem{2020Galax...8...25R}
Javier {Rico}.
\newblock {Gamma-Ray Dark Matter Searches in Milky Way Satellites{\textemdash}A Comparative Review of Data Analysis Methods and Current Results}.
\newblock {\em Galaxies}, 8(1):25, March 2020.

\bibitem{2016JCAP...02..039M}
M.~L. Ahnen et~al.
\newblock {Limits to Dark Matter Annihilation Cross-Section from a Combined Analysis of MAGIC and Fermi-LAT Observations of Dwarf Satellite Galaxies}.
\newblock {\em JCAP}, 02:039, 2016.

\bibitem{2017arXiv170508103I}
M.~G. Aartsen et~al.
\newblock {Search for Neutrinos from Dark Matter Self-Annihilations in the center of the Milky Way with 3 years of IceCube/DeepCore}.
\newblock {\em Eur. Phys. J. C}, 77(9):627, 2017.

\bibitem{2015arXiv150304858T}
K.~Choi et~al.
\newblock {Search for neutrinos from annihilation of captured low-mass dark matter particles in the Sun by Super-Kamiokande}.
\newblock {\em Phys. Rev. Lett.}, 114(14):141301, 2015.

\bibitem{2018PhRvL.120o1301D}
N.~Du et~al.
\newblock {A Search for Invisible Axion Dark Matter with the Axion Dark Matter Experiment}.
\newblock {\em Phys. Rev. Lett.}, 120(15):151301, 2018.

\bibitem{2015ARNPS..65..485G}
Peter~W. {Graham}, Igor~G. {Irastorza}, Steven~K. {Lamoreaux}, Axel {Lindner}, and Karl~A. {van Bibber}.
\newblock {Experimental Searches for the Axion and Axion-Like Particles}.
\newblock {\em Annual Review of Nuclear and Particle Science}, 65:485--514, October 2015.

\bibitem{2020arXiv200610735K}
Kristjan {Kannike}, Martti {Raidal}, Hardi {Veerm{\"a}e}, Alessandro {Strumia}, and Daniele {Teresi}.
\newblock {Dark Matter and the XENON1T electron recoil excess}.
\newblock {\em arXiv e-prints}, page arXiv:2006.10735, June 2020.

\bibitem{2020arXiv200612488B}
Jatan {Buch}, Manuel~A. {Buen-Abad}, JiJi {Fan}, and John~Shing {Chau Leung}.
\newblock {Galactic Origin of Relativistic Bosons and XENON1T Excess}.
\newblock {\em arXiv e-prints}, page arXiv:2006.12488, June 2020.

\bibitem{Steigman:1984ac}
Gary Steigman and Michael~S. Turner.
\newblock {Cosmological Constraints on the Properties of Weakly Interacting Massive Particles}.
\newblock {\em Nucl. Phys. B}, 253:375--386, 1985.

\bibitem{Preskill:1982cy}
John Preskill, Mark~B. Wise, and Frank Wilczek.
\newblock {Cosmology of the Invisible Axion}.
\newblock {\em Phys. Lett.}, B120:127--132, 1983.
\newblock [,URL(1982)].

\bibitem{Abbott:1982af}
L.~F. Abbott and P.~Sikivie.
\newblock {A Cosmological Bound on the Invisible Axion}.
\newblock {\em Phys. Lett.}, B120:133--136, 1983.
\newblock [,URL(1982)].

\bibitem{Dine:1982ah}
Michael Dine and Willy Fischler.
\newblock {The Not So Harmless Axion}.
\newblock {\em Phys. Lett.}, B120:137--141, 1983.
\newblock [,URL(1982)].

\bibitem{Peccei:1977hh}
R.~D. Peccei and Helen~R. Quinn.
\newblock {CP Conservation in the Presence of Instantons}.
\newblock {\em Phys. Rev. Lett.}, 38:1440--1443, 1977.

\bibitem{Wilczek:1977pj}
Frank Wilczek.
\newblock {Problem of Strong $P$ and $T$ Invariance in the Presence of Instantons}.
\newblock {\em Phys. Rev. Lett.}, 40:279--282, 1978.

\bibitem{Weinberg:1977ma}
Steven Weinberg.
\newblock {A New Light Boson?}
\newblock {\em Phys. Rev. Lett.}, 40:223--226, 1978.

\bibitem{DiLuzio:2020wdo}
Luca Di~Luzio, Maurizio Giannotti, Enrico Nardi, and Luca Visinelli.
\newblock {The landscape of QCD axion models}.
\newblock {\em Phys. Rept.}, 870:1--117, 2020.

\bibitem{Turner:1983he}
Michael~S. Turner.
\newblock {Coherent Scalar Field Oscillations in an Expanding Universe}.
\newblock {\em Phys. Rev. D}, 28:1243, 1983.

\bibitem{Press:1989id}
William~H. Press, Barbara~S. Ryden, and David~N. Spergel.
\newblock {Single Mechanism for Generating Large Scale Structure and Providing Dark Missing Matter}.
\newblock {\em Phys. Rev. Lett.}, 64:1084, 1990.

\bibitem{Sin:1992bg}
Sang-Jin Sin.
\newblock {Late time cosmological phase transition and galactic halo as Bose liquid}.
\newblock {\em Phys. Rev.}, D50:3650--3654, 1994.

\bibitem{Hu:2000ke}
Wayne Hu, Rennan Barkana, and Andrei Gruzinov.
\newblock {Cold and fuzzy dark matter}.
\newblock {\em Phys. Rev. Lett.}, 85:1158--1161, 2000.

\bibitem{Goodman:2000tg}
Jeremy Goodman.
\newblock {Repulsive dark matter}.
\newblock {\em New Astron.}, 5:103, 2000.

\bibitem{Peebles:2000yy}
P.~J.~E. Peebles.
\newblock {Fluid dark matter}.
\newblock {\em Astrophys. J. Lett.}, 534:L127, 2000.

\bibitem{Amendola:2005ad}
Luca Amendola and Riccardo Barbieri.
\newblock {Dark matter from an ultra-light pseudo-Goldsone-boson}.
\newblock {\em Phys. Lett. B}, 642:192--196, 2006.

\bibitem{Schive:2014dra}
Hsi-Yu Schive, Tzihong Chiueh, and Tom Broadhurst.
\newblock {Cosmic Structure as the Quantum Interference of a Coherent Dark Wave}.
\newblock {\em Nature Phys.}, 10:496--499, 2014.

\bibitem{Chavanis:2011zi}
Pierre-Henri Chavanis.
\newblock {Mass-radius relation of Newtonian self-gravitating Bose-Einstein condensates with short-range interactions: I. Analytical results}.
\newblock {\em Phys. Rev. D}, 84:043531, 2011.

\bibitem{Chavanis:2011zm}
P.~H. Chavanis and L.~Delfini.
\newblock {Mass-radius relation of Newtonian self-gravitating Bose-Einstein condensates with short-range interactions: II. Numerical results}.
\newblock {\em Phys. Rev. D}, 84:043532, 2011.

\bibitem{Alexander:2016glq}
Stephon Alexander and Sam Cormack.
\newblock {Gravitationally bound BCS state as dark matter}.
\newblock {\em JCAP}, 1704(04):005, 2017.

\bibitem{Alexander:2018fjp}
Stephon Alexander, Evan McDonough, and David~N. Spergel.
\newblock {Chiral Gravitational Waves and Baryon Superfluid Dark Matter}.
\newblock {\em JCAP}, 1805(05):003, 2018.

\bibitem{Alexander:2024nvi}
Stephon Alexander, Tucker Manton, and Evan McDonough.
\newblock {Field theory axiverse}.
\newblock {\em Phys. Rev. D}, 109(11):116019, 2024.

\bibitem{Hui:2016ltb}
Lam Hui, Jeremiah~P. Ostriker, Scott Tremaine, and Edward Witten.
\newblock {Ultralight scalars as cosmological dark matter}.
\newblock {\em Phys. Rev.}, D95(4):043541, 2017.

\bibitem{Ferreira:2020fam}
Elisa G.~M. Ferreira.
\newblock {Ultra-light dark matter}.
\newblock {\em Astron. Astrophys. Rev.}, 29(1):7, 2021.

\bibitem{Capanelli:2025nrj}
Christian Capanelli, Wayne Hu, and Evan McDonough.
\newblock {Wave Interference in Self-Interacting Fuzzy Dark Matter}.
\newblock 3 2025.

\bibitem{Rindler-Daller:2011afd}
Tanja Rindler-Daller and Paul~R. Shapiro.
\newblock {Angular Momentum and Vortex Formation in Bose-Einstein-Condensed Cold Dark Matter Haloes}.
\newblock {\em Mon. Not. Roy. Astron. Soc.}, 422:135--161, 2012.

\bibitem{Alexander:2019puy}
Stephon Alexander, Sergei Gleyzer, Evan McDonough, Michael~W. Toomey, and Emanuele Usai.
\newblock {Deep Learning the Morphology of Dark Matter Substructure}.
\newblock {\em Astrophys. J.}, 893:15, 2020.

\bibitem{Alexander:2021zhx}
Stephon Alexander, Christian Capanelli, Elisa G.~M.~Ferreira, and Evan McDonough.
\newblock {Cosmic filament spin from dark matter vortices}.
\newblock {\em Phys. Lett. B}, 833:137298, 2022.

\bibitem{Alexander:2019qsh}
Stephon Alexander, Jason~J. Bramburger, and Evan McDonough.
\newblock {Dark Disk Substructure and Superfluid Dark Matter}.
\newblock {\em Phys. Lett. B}, 797:134871, 2019.

\bibitem{Alexander:2022own}
Stephon Alexander, Heliudson Bernardo, and Michael~W. Toomey.
\newblock {Addressing the Hubble and $S_8$ Tensions with a Kinetically Mixed Dark Sector}.
\newblock 7 2022.

\bibitem{Bernardo:2022ztc}
Heliudson Bernardo, Robert Brandenberger, and J\"urg Fr\"ohlich.
\newblock {Towards a dark sector model from string theory}.
\newblock {\em JCAP}, 09:040, 2022.

\bibitem{Brax:2023qyp}
Philippe Brax, C.~P. Burgess, and F.~Quevedo.
\newblock {Axio-Chameleons: a novel string-friendly multi-field screening mechanism}.
\newblock {\em JCAP}, 03:015, 2024.

\bibitem{DESI:2025zgx}
M.~Abdul~Karim et~al.
\newblock {DESI DR2 Results II: Measurements of Baryon Acoustic Oscillations and Cosmological Constraints}.
\newblock 3 2025.

\bibitem{DESI:2024mwx}
A.~G. Adame et~al.
\newblock {DESI 2024 VI: cosmological constraints from the measurements of baryon acoustic oscillations}.
\newblock {\em JCAP}, 02:021, 2025.

\bibitem{Smith:2024ibv}
Adam Smith, Maria Mylova, Philippe Brax, Carsten van~de Bruck, C.~P. Burgess, and Anne-Christine Davis.
\newblock {A Minimal Axio-dilaton Dark Sector}.
\newblock 10 2024.

\bibitem{Burgess:2021obw}
C.~P. Burgess, Danielle Dineen, and F.~Quevedo.
\newblock {Yoga Dark Energy: natural relaxation and other dark implications of a supersymmetric gravity sector}.
\newblock {\em JCAP}, 03(03):064, 2022.

\bibitem{Guth:2014hsa}
Alan~H. Guth, Mark~P. Hertzberg, and C.~Prescod-Weinstein.
\newblock {Do Dark Matter Axions Form a Condensate with Long-Range Correlation?}
\newblock {\em Phys. Rev. D}, 92(10):103513, 2015.

\bibitem{Caldwell:1997ii}
R.~R. Caldwell, Rahul Dave, and Paul~J. Steinhardt.
\newblock {Cosmological imprint of an energy component with general equation of state}.
\newblock {\em Phys. Rev. Lett.}, 80:1582--1585, 1998.

\bibitem{Marsh:2010wq}
David J.~E. Marsh and Pedro~G. Ferreira.
\newblock {Ultra-Light Scalar Fields and the Growth of Structure in the Universe}.
\newblock {\em Phys. Rev. D}, 82:103528, 2010.

\bibitem{Allali:2021azp}
Itamar~J. Allali, Mark~P. Hertzberg, and Fabrizio Rompineve.
\newblock {Dark sector to restore cosmological concordance}.
\newblock {\em Phys. Rev. D}, 104(8):L081303, 2021.

\bibitem{Rogers:2020ltq}
Keir~K. Rogers and Hiranya~V. Peiris.
\newblock {Strong Bound on Canonical Ultralight Axion Dark Matter from the Lyman-Alpha Forest}.
\newblock {\em Phys. Rev. Lett.}, 126(7):071302, 2021.

\bibitem{Goldstein:2022pxu}
Isabelle~S. Goldstein, Savvas~M. Koushiappas, and Matthew~G. Walker.
\newblock {Viability of ultralight bosonic dark matter in dwarf galaxies}.
\newblock {\em Phys. Rev. D}, 106(6):063010, 2022.

\bibitem{DES:2020fxi}
E.~O. Nadler et~al.
\newblock {Milky Way Satellite Census. III. Constraints on Dark Matter Properties from Observations of Milky Way Satellite Galaxies}.
\newblock {\em Phys. Rev. Lett.}, 126:091101, 2021.

\bibitem{Ivanov:2023yla}
Mikhail~M. Ivanov.
\newblock {Lyman alpha forest power spectrum in effective field theory}.
\newblock {\em Phys. Rev. D}, 109(2):023507, 2024.

\bibitem{Ivanov:2024jtl}
Mikhail~M. Ivanov, Michael~W. Toomey, and Naim~G\"oksel Kara\c{c}ayl\i{}.
\newblock {Fundamental Physics with the Lyman-Alpha Forest: Constraints on the Growth of Structure and Neutrino Masses from SDSS with Effective Field Theory}.
\newblock {\em Phys. Rev. Lett.}, 134(9):091001, 2025.

\bibitem{Qin:2022xho}
Wenzer Qin, Katelin Schutz, Aaron Smith, Enrico Garaldi, Rahul Kannan, Tracy~R. Slatyer, and Mark Vogelsberger.
\newblock {Effective bias expansion for 21-cm cosmology in redshift space}.
\newblock {\em Phys. Rev. D}, 106(12):123506, 2022.

\bibitem{Dine:1985he}
Michael Dine and Nathan Seiberg.
\newblock {Is the Superstring Weakly Coupled?}
\newblock {\em Phys. Lett. B}, 162:299--302, 1985.

\bibitem{Alexander:2019rsc}
Stephon Alexander and Evan McDonough.
\newblock {Axion-Dilaton Destabilization and the Hubble Tension}.
\newblock {\em Phys. Lett. B}, 797:134830, 2019.

\bibitem{Wang:2025vfb}
Yun Wang and Katherine Freese.
\newblock {Model-Independent Dark Energy Measurements from DESI DR2 and Planck 2015 Data}.
\newblock 5 2025.

\bibitem{Efstathiou:2025tie}
George Efstathiou.
\newblock {Baryon Acoustic Oscillations from a Different Angle}.
\newblock 5 2025.

\bibitem{Goldstein:2025lvl}
Isabelle~S. Goldstein and Louis~E. Strigari.
\newblock {Kinematics of the Sagittarius Dwarf Spheroidal core: A 5D Analysis for a 6D Methodology with Gaia DR3}.
\newblock 1 2025.

\bibitem{Weltman:2018zrl}
A.~Weltman et~al.
\newblock {Fundamental physics with the Square Kilometre Array}.
\newblock {\em Publ. Astron. Soc. Austral.}, 37:e002, 2020.

\bibitem{DESI:2025zpo}
M.~Abdul~Karim et~al.
\newblock {DESI DR2 Results I: Baryon Acoustic Oscillations from the Lyman Alpha Forest}.
\newblock 3 2025.

\bibitem{Moresco:2022phi}
Michele Moresco et~al.
\newblock {Unveiling the Universe with emerging cosmological probes}.
\newblock {\em Living Rev. Rel.}, 25(1):6, 2022.

\end{thebibliography}

\end{document}